\documentclass[10pt,final,journal,twocolumn]{IEEEtran}
\ifCLASSINFOpdf
\else
\fi
\usepackage{cite}
\usepackage[cmex10]{amsmath}
\usepackage{amsthm}
\usepackage{algorithmic}
\usepackage{stfloats}
\usepackage{multicol,multienum}
\usepackage[dvips]{graphicx}
\usepackage{color}

\begin{document}

\title{Dynamic Spectrum Access in Time-varying Environment: Distributed Learning Beyond Expectation Optimization}

\author{
Yuhua~Xu,~\IEEEmembership{Member,~IEEE,}
Jinlong~Wang,~\IEEEmembership{Senior Member,~IEEE,}
Qihui~Wu,~\IEEEmembership{Senior Member,~IEEE,}
Jianchao~Zheng, ~\IEEEmembership{ Member,~IEEE,}
Liang Shen, and Alagan~Anpalagan,~\IEEEmembership{Senior Member,~IEEE}

\thanks{A part of this work has been accepted by IEEE International Conference on Communications (ICC'17). This work was supported by Natural Science Foundation for Distinguished Young Scholars of Jiangsu Province under Grant No. BK20160034, the National Science Foundation of China under Grant No. 61671473, No. 61401508 and No. 61631020, and in part by the Open Research Foundation of Science and Technology on Communication Networks Laboratory. }

\thanks{Yuhua~Xu is with the College of Communications Engineering, PLA University of Science and Technology, Nanjing, China. He is also with the Science and Technology on Communication Networks Laboratory, Shijiazhuang, China. (yuhuaenator@gmail.com). }

\thanks{Jinlong~Wang, Jianchao Zheng and Liang Shen are with the College of Communications Engineering, PLA University of Science and Technology, Nanjing, China (wjl543@sina.com, longxingren.zjc.s@163.com, ShenLiang671104@sina.com). }

\thanks{Qihui Wu is  with the College of Electronic and Information Engineering,
Nanjing University of Aeronautics and Astronautics, Nanjing, China (e-mail: wuqihui2014@sina.com).}

\thanks{Alagan~Anpalagan is with the Department of Electrical and Computer Engineering, Ryerson University, Toronto, Canada (alagan@ee.ryerson.ca).}
}

%
%

\IEEEpeerreviewmaketitle
\maketitle
\begin{abstract}
This article investigates the problem of dynamic spectrum access for canonical wireless networks, in which the channel states are time-varying. In the most existing work, the commonly used optimization objective is to maximize the expectation of a certain metric (e.g., throughput or achievable rate). However, it is realized that expectation alone is not enough since some applications are sensitive to  fluctuations. Effective capacity is a promising metric for time-varying service process since it characterizes the  packet delay violating probability (regarded as an important statistical QoS index),  by taking into account not only the expectation but also other high-order statistic. Therefore, we formulate the interactions among the users in the time-varying environment as a non-cooperative game, in which the utility function is defined as the achieved effective capacity. We prove that it is an ordinal potential game which has at least one pure strategy Nash equilibrium. Based on an approximated utility function, we propose a multi-agent learning algorithm which is proved to achieve stable solutions with \emph{dynamic} and \emph{incomplete} information constraints. The convergence of the proposed learning algorithm is verified by simulation results. Also, it is shown that the proposed multi-agent learning algorithm achieves satisfactory performance.
\end{abstract}


%
\IEEEpeerreviewmaketitle

\section{Introduction}
\IEEEPARstart {D}{ynamic} spectrum access (DSA) has been regarded as one of the most important technology for future wireless networks since it provides flexible and efficient spectrum usage.
With the significant advances in cognitive radios in the last decade \cite{attention1,attention3} , DSA can be implemented in more intelligent and smart manners \cite{Decision_Theoretic_CR,Spectrum_Decision_CR}.
Generally, there are two main application scenarios \cite{DSA_CR}: \emph{open-access}, in which all users are equal to access the spectrum,  and \emph{primary-secondary access}, in which the spectrum is owned by the primary users and can be used by the secondary users when it is idle. For decision-making, it has been shown that the methodologies for the two scenarios are mostly overlapped \cite{Decision_Theoretic_CR}.

A number of existing studies, e.g., \cite{static_selection1,static_selection2,static_selection3,static_selection4,static_selection5,static_selection6}, have considered intelligent spectrum access  for static wireless networks in which the channel states remain unchanged during the selection procedure.  However, it has been realized that although the assumption of static channel leads to mathematical tractability, it is not generally true since the spectrum are always time-varying in wireless environment \cite{dynamic_selection1,dynamic_selection2,dynamic_selection3}.
To track the channel dynamics, an instinctive approach is to reiterate the selection algorithms in each quasi-static period. This method, however, is off-line, costly and inefficient, and is even not feasible for fast-varying channels. Thus, it is timely important to develop  on-line intelligent channel selection algorithms for dynamic wireless networks.

In this article, we consider a dynamic wireless canonical networks, in which the channel states are time-varying and there is no information exchange among the users. In a few existing researches for dynamic networks with time-varying channels, e.g.,\cite{dynamic_selection1,dynamic_selection2,dynamic_selection3,MAB1}, the commonly used optimization objective is to maximize the expectation of a certain metric, e.g., the expected throughput. However, only considering the expectation is not enough for practical applications. For example, in real-time multimedia applications, higher expected transmission rate as well as lower fluctuation are desirable, which implies that not only the expectation but also other statistic, e.g., the variance,  should be taken into account for dynamic wireless networks. A promising metric is the effective capacity, which is defined as the maximum packet arrival rate that a time-varying service process can support while a statistical quality-of-service (QoS) constraint on delay violating probability can be met \cite{effective_capacity1}. Mathematically, effective capacity takes  into account the expectation and all other statistics \cite{Learning_book}; further, it degrades  the expectation if the statistical QoS index is sufficiently small. Therefore, we use effective capacity as the optimization metric in this article\footnote{It should be pointed out that the main focus of this paper is to consider both expectation and other statistic in dynamic wireless networks. Thus, except for the used effective capacity, other forms of optimization metric can also be used. We will explain this more specific later.}.

The considered  DSA network encounters \emph{dynamic} and \emph{incomplete} information constraints for the decision procedure. Specifically, the channel states are not deterministic at each slot and change from slot to slot, and a user can only monitor its chosen channel and know nothing about other users. Furthermore, the introduction of effective capacity into dynamic cognitive radio networks leads to additional challenges. In comparison, the expectation admits additive property in the time domain while the effective capacity does not. In particular, an expected value can be obtained by cumulatively averaging the random payoffs in a long  period. However, effective capacity does not admit the additive property due to its nonlinearity. Thus, the task of designing effective-capacity oriented intelligent channel selection approaches for multiple users with the  dynamic and incomplete information constraints remains unsolved and is challenging.

Since the decisions of the users are interactive, we formulate the problem of dynamic spectrum access in time-varying environment as a non-cooperative game, in which  utility function is defined as the effective capacity. We prove that it is an ordinal potential game which has at least one pure strategy Nash equilibrium (NE). Due to the  dynamic and incomplete information constraints, existing game-theoretic algorithms, e.g., the best response \cite{potential_game}, fictitious play \cite{game_book}, spatial adaptive play \cite{static_selection1} and regret learning \cite{static_selection2}, can not be applied to the considered dynamic networks. The reason is that they are originally designed for static systems with complete information. It is known that users in cognitive radios are able to observe the environment, learn from history experiences, and make intelligent decisions \cite{attention3}. Following the
CODIPAS learning techniques \cite{Learning_book} (COmbined fully DIstributed PAyoff and Strategy),
 we  propose a multi-agent learning algorithm to achieve the Nash equilibria of the formulated dynamic spectrum access game in time-varying environment. To summarize, the main contributions of this article are:
\begin{enumerate}
  \item We formulate the problem of dynamic spectrum access in time-varying environment  as a non-cooperative game, in which the utility function of each user is defined as the effective capacity characterized by a statistical QoS index. In particular, the utility function takes into account not only the expectation of the achievable transmission rate but also other statistic. We prove that the game is an ordinal potential game and hence has at least one pure strategy NE point.
  \item Based on an approximated utility function, we propose a multi-agent learning algorithm to achieve the pure strategy NE points of the game with unknown, dynamic and incomplete information constraints. The proposed algorithm is fully distributed and autonomous, since it only relies on the individual information of a user and does not need information about other players. Simulation results show that the proposed learning algorithm achieves satisfactory performance.
\end{enumerate}

Note that there are some previous work which also considered effective capacity in dynamic spectrum access/cognitive radio networks, e.g., \cite{ effective_capacity3,effective_capacity4,effective_capacity5,effective_capacity6}. The main differences in methodology are: i) most existing studies
considered optimization of effective capacity in a centralized manner, while we consider it in a distributed manner,
 ii) we consider the interactions among multiple users and propose a multi-agent learning algorithm to achieve stable solutions,
and iii) the effective capacity can not be obtained by cumulatively averaging the random payoffs in a long  period due to its nonlinearity, which brings new challenges for the learning solutions.

Also, it should be pointed out that the presented game model is motivated by the risk-sensitive game proposed in \cite{Learning_book}, which admits the same utility function. The key differences in this paper are: (i) the effective capacity has physical meaning in wireless communications, i.e., it implies statistical QoS provisioning,  and (ii) we show that the dynamic spectrum access game with effective capacity optimization is an ordinal potential game.

The rest of the article  is organized as follows. In Section II, we give a brief review of related work. In Section III, we present the system model and formulate the problem. In Section IV, we present the dynamic spectrum access game and investigate the properties of its NE, and propose a multi-agent learning algorithm for achieving stable solutions. In Section V, simulation results are presented. Finally, we present discussion and draw conclusion in Section VI.

\section{Related Work}
The problem of dynamic spectrum access in both open-access and primary-secondary access scenarios  has been extensively investigated in the context of  cognitive radio, e.g., \cite{static_selection1,static_selection2,static_selection3,static_selection4,static_selection5,static_selection6,static_selection7,static_selection8,static_selection10}.
These work mainly focused on static  networks, in which the channel states remain unchanged during the learning and decision procedure. However, it has been realized that the assumption of static channel is not always true in practice. Recently, the problem dynamic spectrum access with varying channel states began to draw attention, using e.g.,
Markovian decision process (MDP) \cite{dynamic_selection2}, online learning algorithms for multi-armed bandit (MAB) problems \cite{MAB1}, and game-theoretic learning \cite{dynamic_selection1,dynamic_selection3}. The commonly used optimization metric in these work is to maximize the expected achievable transmission rate, which does not consider the QoS requirement in the packet delay. In addition, the algorithms in MDP and MAB models are mainly for scenarios with single user. Compared with those existing studies, this work is differentiated in that a statistical QoS requirement in packet delay is considered for a multiuser DSA network with time-varying channels.

It is noted that multi-agent learning algorithms for game-theoretic solutions in wireless networks have been an active topic. Specifically, stochastic learning automata \cite{Sastry94} based algorithms for wireless communications can be found in the literature, e.g., distributed channel selection for opportunistic spectrum access \cite{dynamic_selection1,dynamic_selection3}, distributed power control \cite{SLA}, precoding selection for MIMO systems \cite{SLA2}, spectrum management \cite{Q_learning_AAMAS10} and cooperative coordination design \cite{Q_learning_JSAC13} for cognitive radio networks. Furthermore, Q-learning based dynamic spectrum access was reported in \cite{Li_EURASIP10, Li09,Li10},
various combined fully distributed payoff and strategy-reinforcement learning algorithms for 4G heterogeneous networks were studied in \cite{Learning_Cost}, a trial-and-error learning approach for self-organization in decentralized networks was studied in \cite{trial_learning}, and several variations of logit-learning algorithms were studied in \cite{static_selection5,static_selection6}. In methodology, all of the above mentioned algorithms are originally designed for maximizing the expectation and hence can not be applied. We consider a new optimization metric that takes into account not only the expectation but also  other high-order moments.

The most related work is \cite{effective_capacity7}, in which a game-theoretic optimization approach for effective capacity  in cognitive femtocells was studied. The key difference in this work is that we focus on
formulating the game model as well as designing multi-agent learning with  dynamic and incomplete information constraints. Nevertheless, the authors of \cite{effective_capacity7} only focused on  game formulation and analysis. Another related work is \cite{satisfaction_equilibrium}, in which a satisfaction equilibrium approach is proposed for  QoS provisioning in decentralized networks. Note that a part of this work with some preliminary results can be found in \cite{ICC17_Xu}.

Note that NE may be inefficient due to its inherent non-cooperative nature. There are some other solutions beyond NE to improve the efficiency, e.g., pricing \cite{pricing}, auction \cite{auction}, Nash bargaining \cite{bargaining}, and coalitional games \cite{Coalitional1,Coalitional2}. The key difference in this paper is that the proposed solution does not need information exchange while these solutions need information exchange among users, which may cause heavy communication overhead.

\section{System Models and Problem Formulation}
\subsection{System model}
We consider a distributed canonical network consisting of $N$ users and $M$  channels. A user in canonical networks is a collection of multiple entities with intra-communications and there is a heading managing the whole community \cite{Babadi10}. Examples of users in canonical networks given by, e.g., a WLAN access point with the serving clients \cite{Cao08}, a small cell base station with its mobile terminals \cite{Xu_magazine},  and a cluster head with its belonged members. For presentation, denote the user set as $\mathcal{N}$, i.e., $\mathcal{N}=\{1,\dots,N\}$, and the channel set as $\mathcal{M}$, i.e., $\mathcal{M}=\{1,\dots,M\}$. Due to fading in wireless environment, the transmission rate of each channel is always time-varying. To capture the rate fluctuations, the finite rate channel model is applied \cite{finite_rate_model}. In particular, the rate set of  channel $m$ is denoted as $\mathcal{S}_m=\{s_{m1}, s_{m2},\ldots, s_{mK}\}$, where $s_{mk}$ indicates that the channel can support certain transmission rate (packets/slot). The corresponding rate-state probabilities are given by $\Pi_m=\{\pi_{m1},\ldots, \pi_{mK}\}$ and the expected transmission rate of channel $m$ is given by $\bar s_m=\sum\nolimits_k \pi_{mk} s_{mk}$. The users do not know the rate distribution of the channels.

We assume that time is divided into slots with equal length and the transmission rate of each channel is block-fixed in a slot and changes randomly in the next slot. Specifically, the achievable transmission rate of channel $m$ for user $n$ in slot $i$ is denoted as $r_{nm}(i)$, which is randomly chosen from the  rate set $\mathcal{S}_m$. We consider heterogeneous spectrum in this article, i.e.,
 the transmission rate set and the corresponding probability set vary from channel to channel\footnote{The feature of heterogeneous spectrum is caused by the flexible spectrum usage pattern in current wireless communication systems.  Examples are given by: (i) in cognitive radio networks, the channels are occupied by the primary users with different probabilities and (ii) in heterogeneous networks, the channels belong to different networks have different rate sets.}.

The task of each user is to choose an appropriate channel to access. Without loss of generality, we assume that the number of users is larger  than that of the channels, i.e., $N>M$. When more than one user chooses the same channel, they share the channel using some multiple access mechanisms, e.g., TDMA or CSMA. There is no central controller and no information exchange among the users, which means that the users should choose appropriate channels through learning and adjusting.

 Denote $a_n$ as the chosen channel  of user $n$, i.e., $a_n \in \mathcal{M}$. In the following, we analyze the achievable transmission rates of the users for different  multiple access mechanisms\cite{ICC17_Xu}:
\begin{enumerate}
  \item  If perfect TDMA is applied to resolve contention among the users,  the instantaneous achievable transmission rate of user $n$ is determined as follows: all the  users  share the channel equally. Thus,
  the instantaneous achievable rate of user $n$ is as follows:
   \begin{equation}
 \label{eq:random_rate_TDMA}
   r_n(t)=\frac{s_{a_n}(t)}{1+\sum\limits_{i\in{\mathcal{N}, n\neq i }} I(a_n,a_{i})},
 \end{equation}
  where $s_{a_n}(t)$ is the instantaneous transmission rate of channel $a_n$ in time $t$, and  $I(a_n,a_{n'})$ is the following indicator function:
  \begin{equation}
 \label{eq:indicator}
  I(a_n,a_{n'}) =\left\{ \begin{array}{l}
 1,\;\;\;\;\;\;\;\;\;\;\; a_n=a_{n'} \\
 0,\;\;\;\;\;\;\;\;\;\;\; a_n\neq a_{n'}\\
 \end{array} \right.
 \end{equation}

  \item If perfect CSMA is applied,  the instantaneous achievable transmission rate of user $n$ is determined as follows: only a user can transmit successfully and all other users on the same channel must stay silent. Thus, the instantaneous achievable rate of user $n$ is as follows:

   \begin{equation}
 \label{eq:random_rate_CSMA}
   r_n(t) =\left\{ \begin{array}{l}
  s_{a_n}(t), \;\;\;\; \text{w}.\text{p}. \;\;  \Big( \frac{1}{1+\sum\limits_{i\in{\mathcal{N}, n\neq i }} I(a_n,a_{i})} \Big)\\
   0,\;\;\;\;\;\;\text{w}.\text{p}. \;\; \Big(1-\frac{1}{1+\sum\limits_{l\in{\mathcal{N}, n\neq l }} I(a_n,a_{l})}\Big), \\
 \end{array} \right.
 \end{equation}
\end{enumerate}


\subsection{Preliminary of effective capacity}
 Since the channel transmission rate are time-varying, one candidate optimization matric is to maximize the expected transmission rate of user $n$, i.e., $\max \text E [{r_n(t)}]$. It is noted that such an objective is not enough since the rate fluctuation may cause severe delay-bound violating probability whereas the expected rate cannot reflect this event. To study the effect of time-varying transmission rate, one would take into account not only the expectation but  also the variance
 and other higher-order the moments. Among all possible solutions, the theory of effective capacity of time-varying service process is a promising
 approach. Therefore, we use effective capacity to study the problem of opportunistic channel access in heterogeneous spectrum.

Using the large deviations theory \cite{Large_deviation}, it was shown in \cite{effective_capacity} that for a dynamic queuing system with stationary arrival and service processes, the probability that the stationary queue length $L(t)$ is large than a threshold $l$ is given by:
    \begin{equation}
    \label{eq: queue_violating}
 \lim_{l\rightarrow \infty} \Big[ \frac{\log \Pr\{L(t)>l\}}{l}\Big]=-\theta,
 \end{equation}
  where $\theta$ serves as the exponential decay rate tail distribution of the stationary queue length. Therefore, for sufficiently large $l$, the queue length violating probability can be approximated by $\Pr\{L(t)>l\}\approx e^{-\theta l}$. It is shown that  larger $\theta$ corresponds to strict QoS requirement while small $\theta$ implies  loose QoS requirement. Furthermore,  for a stationary traffic with fixed arrival rate $\alpha$, the  delay-bound violating probability and the length-bound violating probability is related by:
  \begin{equation}
  \Pr \{D(t)>d\}\le c \sqrt{\Pr \{L(t)>l\}},
 \end{equation}
 where $c$ is some positive constant and $l= \alpha d$. From the above analysis, it is seen that both the queue length violating probability and delay-bound violating probability are determined by the exponential decay rate $\theta$, which specifies the QoS requirement. Thus, we will pay attention to $\theta$ in this article.

 For a time-varying service process with independent and identical distribution (i.i.d.), the \emph{effective capacity} is defined as follows \cite{effective_capacity}:
 \begin{equation}
 \label{eq:effective_capacity_definition}
C(\theta)=- \frac{1}{\theta} \log \big( \text{E} [ e^{-\theta x(t) }] \big),
 \end{equation}
where $x(t)$ is the time-varying service process, and $\theta$ is the statistical QoS index as specified by (\ref{eq: queue_violating}). The properties of effective capacity is analyzed as follows \cite{Learning_book}:
\begin{itemize}
\item For a given time-varying service, it is a decreasing function with respect to $\theta$, i.e.,
   \begin{equation}
 \label{eq:property1}
  0<\theta_2<\theta_1 \Rightarrow C(\theta_1) < C(\theta_2).
 \end{equation}
\item For  each $\theta>0$, the effective capacity is always less than the expected capacity, i.e.,
   \begin{equation}
 \label{eq:property2}
C(\theta)< \text{E} [  x(t) ], \forall \theta>0,
 \end{equation}
 which can be proved by Jensen's inequality \cite{inequality}.

  \item As $\theta$ approaches zero, the effective capacity degrades to the expected capacity, i.e.,
   \begin{equation}
 \label{eq:property3}
\lim_{\theta \rightarrow 0} C(\theta)= \text{E} [  x(t) ].
 \end{equation}
  \item If $\theta$ is sufficiently small, by performing Taylor expansion, we have:
    \begin{equation}
 \label{eq:property4}
C(\theta)=\text{E} [  x(t) ]-\frac{\theta}{2}\text{var}[x(t)]+\text{o}(\theta),
 \end{equation}
 where $\text{var}[x(t)]$ is the variance of $x(t)$, and $\text{o}(\theta)$ is the  infinitely small quantity of higher order.
\end{itemize}

From (\ref{eq:property1}) to (\ref{eq:property4}), it is seen that the effective capacity takes into account not only the expectation but also other moments (including the variance and other high-order moments) to capture the fluctuation in the time-varying service rate.

\subsection{Problem formulation}
For the considered dynamic spectrum access system, we use the effective capacity as the optimization metric. Specifically, denote $\theta_n$ as the statistical QoS index of user $n$, then the achievable effective capacity of user $n$ is given by
\begin{equation}
 \label{eq:effective_capacity_definition}
C_n(a_n, a_{-n}, \theta_n)=- \frac{1}{\theta_n} \log \big( \text{E} [ e^{-\theta_n r_n(i) }] \big),
 \end{equation}
where $r_n(i)$ is the instantaneous transmission rate as specified by (\ref{eq:random_rate_TDMA}) or (\ref{eq:random_rate_CSMA}) , and $a_{-n}$ is the channel selection profile of all the users except user $n$.

For each user, the optimization objective is to choose a channel to maximize the effective capacity\footnote{Since the main concern of this paper is to consider both expectation and other-order statistic for dynamic OSA networks, other forms of optimization metric can also be used. For example, one may use the following  objective:
 \[O_1=\alpha_1\text{E} [  x(t) ]-\alpha_2\text{var}[x(t)],\]
where $\alpha_1$ and $\alpha_2$ are the weighted coefficients determined by the specific practical applications. The reasons for using  effective capacity as the optimization goal in this paper are twofold: (i) effective capacity takes into both expectation and other statistic into account, and (ii) it has physical meanings related to QoS provisioning for time-varying OSA networks. }.
It has been pointed out that information is key to decision-making problems \cite{Decision_Theoretic_CR}. For the considered dynamic spectrum access with statistical QoS provisioning,  the information constraints  can be summarized as follows:
\begin{itemize}
  \item \textbf{Dynamic:} the instantaneous channel transmission  rates are not deterministic, and the event of successfully accessing a channel in a slot is random. Furthermore, the instantaneous channel transmission  rate is time-varying.
  \item \textbf{Incomplete:} the rate-state probabilities of each channel are unknown to the users, and a user does not know the QoS index of other users. Moreover, there is no information exchange among the users.
\end{itemize}

 Due to the above \emph{dynamic} and \emph{incomplete} information constraints, it is challenging to achieve desirable solutions even in a centralized manner, not to mention in an autonomous and distributed manner. Learning, which is core of cognitive radios \cite{attention1}, would achieve satisfactory performance in complex and dynamic environment. In the following, we propose a multi-agent learning approach to solve this problem.

\section{Multi-agent Learning Approach}
Since  there is no central controller, the users behave autonomously and selfishly, i.e., each user optimizes its individual effective capacity. In addition, there is no information exchange between the users, which means that cooperation is not feasible in this scenario. This motivates us to formulate a non-cooperative game to capture the interactions among users.  The properties of the formulated game are investigated. However, due to the  dynamic and incomplete information constraints, most existing game-theoretic algorithms can not be applied. Therefore, we propose a multi-agent learning approach for the users to achieve desirable solutions autonomously and distributively.

\subsection{Dynamic spectrum access game with QoS provisioning}
The dynamic channel access game with QoS provisioning is denoted as $\mathcal{G}=\{\mathcal{N},\theta_n, \mathcal{A}_n, u_n \}$, where $\mathcal{N}$ is the player (user) set, $\mathcal{A}_n$ is the action space of player $n$, $\theta_n$ is the QoS index of player $n$ and $u_n$ is the utility function of player $n$. The action space of each player is exactly the available channel set, i.e., $\mathcal{A}_n\equiv\mathcal{M}$, $\forall n\in \mathcal{N}$. In this game, the utility function is exactly the achievable effective capacity, i.e.,
\begin{equation}
 \label{eq:utility_function}
 u_n(a_n,a_{-n})=- \frac{1}{\theta_n} \log \big( \text{E} [ e^{-\theta_n r_n(i) }] \big),
 \end{equation}
 In non-cooperative games, each player maximizes its individual utility. Therefore, the proposed dynamic spectrum access game with QoS provisioning  can be expressed as:
 \begin{equation}
 \label{eq:game}
 \mathcal{G}: \;\;\;\;\;\;\;\;\;\;  \max u_n(a_n,a_{-n}), \forall n \in \mathcal{N}
 \end{equation}

 For a channel selection profile $(a_n,a_{-n})$,  denote the set of users choosing channel $m$ as $\mathcal{C}_m$, i.e., $\mathcal{C}_m=\{n\in\mathcal{N}: a_n=m\}$, then the number of users choosing channel $m$ can be expressed as
$c_m(a_n,a_{-n})=|\mathcal{C}_m|$=${1+\sum\limits_{i\in{\mathcal{N}, n\neq i }} I(a_n,a_{i})}$.

\subsection{Analysis of Nash equilibrium (NE)}
In this subsection, we present the concept of Nash equilibrium (NE), which is the most well-known stable solution in non-cooperative game models, and analyze its properties. A channel selection profile $a^{*}=(a^{*}_1,\ldots,a^{*}_N)$ is a pure strategy NE if and only if no player can improve its utility function by deviating  unilaterally \cite{game_book}, i.e.,
 \begin{equation}
 \label{eq:NE_definition}
    u_n(a^{*}_n,a^{*}_{-n}) \ge u_n(a_n,a^{*}_{-n}), \forall n\in \mathcal{N}, \forall a_n \in \mathcal{A}_n
 \end{equation}

To investigate the properties of the formulated  game, we first present the following definitions \cite{potential_game}.

\textbf{Definition 1.} A game is  an exact potential game (EPG) if there exists an exact potential function $\phi_e: {{A}_1} \times  \cdots  \times {{A}_N} \to R$ such that for all $n \in \mathcal{N}$, all $a_n \in \mathcal{A}_n$, and $a'_n \in \mathcal{A}_n$,
 \begin{equation}
 \label{eq:EPG_definition}
    u_n(a_n,a_{-n})-u_n(a'_n,a_{-n}) = \phi_e(a_n,a_{-n})-\phi_e(a'_n,a_{-n})
 \end{equation}
In other words, the change in the utility function caused by an arbitrary unilateral action change of a user is the same with that in the exact potential function.

\textbf{Definition 2.} A game is an ordinal potential game (OPG) if there exists an ordinal potential function $\phi_o: {{A}_1} \times  \cdots  \times {{A}_N} \to R$ such that for all $n \in \mathcal{N}$, all $a_n \in \mathcal{A}_n$, and $a'_n \in \mathcal{A}_n$, the following holds:
 \begin{equation}
 \label{eq:OPG_definition}
  \begin{array}{l}
    u_n(a_n,a_{-n})-u_n(a'_n,a_{-n}) \ge 0 \\
    \;\;\;\;\;\;\;\;\;\;\;\;\;\;\;\;\Leftrightarrow \phi_o(a_n,a_{-n})-\phi_o(a'_n,a_{-n}) \ge 0
  \end{array}
 \end{equation}
 In other words, if the change in the utility function caused by an arbitrary unilateral action change is increasing, the change in the ordinal potential function keeps the same trend.

According to the finite improvement property \cite{potential_game}, both EPG and OPG admits the following two promising features: (i) every EPG (OPG)  has at least one pure strategy Nash equilibrium, and (ii) an action profile that maximizes the exact (ordinal) potential function is also a Nash equilibrium.

\newtheorem{theorem}{Theorem}
\begin{theorem}
\label{tm:potential_game}
The  dynamic spectrum access game with effective capacity serving as the utility function  is an OPG, which has at least one pure strategy Nash equilibrium.
\end{theorem}

\begin{IEEEproof}
For presentation, we consider the scenarios that TDMA is applied. To begin with, we omit the logarithmic term in the utility function in (\ref{eq:utility_function}) and denote:
 \begin{equation}
 \label{eq:utility_v}
 v_n(a_n,a_{-n})= \text{E} [ e^{-\theta_n r_n(i) }].
 \end{equation}

 For an arbitrary player $n \in \mathcal{C}_m$, we have:
 \begin{equation}
 \label{eq:utility_function2}
 v_n(a_n,a_{-n})= \sum\limits_{k=1}^K \pi_{mk}e^{-\theta_n \frac{s_{mk}}{c_m}},
 \end{equation}
 where $s_{mk}$ is the random transmission rate of channel $m$ and $\pi_{mk}$ is the corresponding probability.
 For presentation, denote $v_n^{(k)}(a_n,a_{-n})=\pi_{mk}e^{-\theta_n \frac{s_{mk}}{c_m}}$, $k=1,\ldots,K$, which are a family of functions. Motivated by \emph{Rosenthal's potential function} \cite{Vcking06congestiongames}, we define $\phi_v^{(k)} ({a_n},{a_{ - n}}): A_1 \times \cdots A_N \to R$ as
 \begin{equation}
\label{eq:potential_function_v_k}
\phi_v^{(k)} ({a_n},{a_{ - n}}) = \sum\limits_{m = 1}^M {\sum\limits_{l = 1}^{{c _m(a_n,a_{-n})}} \pi_{mk}e^{-\theta_n \frac{s_{mk}}{l}}},
\end{equation}
and
\begin{equation}
\label{eq:potential_function_v}
\phi_v ({a_n},{a_{ - n}}) =\sum\nolimits_{k=1}^{K}\phi_v^{(k)} ({a_n},{a_{ - n}}).
\end{equation}

Now, suppose that player $n$ unilaterally changes its channel selection from $a_n$ to $a'_n$ (denote $a'_n=m'$ for presentation),   the change in $v_n^{(k)}(a_n,a_{-n})$ caused by this unilateral change can be expressed as:
\begin{equation}
\label{eq:potential_function}
\begin{array}{l}
v_n^{(k)}(a'_n,a_{-n})-v_n^{(k)}(a_n,a_{-n})\\
=\pi_{m'k}e^{-\theta_n \frac{s_{m'k}}{c_{m'}(a'_n,a_{-n})}}-\pi_{mk}e^{-\theta_n \frac{s_{mk}}{c_m(a_n,a_{-n})}}
\end{array}
\end{equation}

\begin{figure*}[tb!]

\begin{equation}
\label{eq:potential_change_v}
\phi_v^{(k)} ({a'_n},{a_{ - n}})-\phi_v^{(k)} ({a_n},{a_{ - n}})=\sum\limits_{m' = 1}^M {\sum\limits_{l = 1}^{{c _{m'}}} \pi_{m'k}e^{-\theta_n \frac{s_{m'k}}{l}}}-\sum\limits_{m = 1}^M {\sum\limits_{l = 1}^{{c_m}} \pi_{mk}e^{-\theta_n \frac{s_{mk}}{l}}}
\end{equation}

\begin{equation}
\label{eq:potential_change_v2}
\begin{array}{l}
\phi_v^{(k)} ({a'_n},{a_{ - n}})-\phi_v^{(k)} ({a_n},{a_{ - n}})= \Bigg( {\sum\limits_{l = 1}^{{c _{m'}(a_n,a_{-n})+1}} \pi_{m'k}e^{-\theta_n \frac{s_{m'k}}{l}}}+{\sum\limits_{l = 1}^{{c_m}(a_n,a_{-n})-1} \pi_{mk}e^{-\theta_n \frac{s_{mk}}{l}}} \Bigg)\\
\;\;\;\;\;\;\;\;\;\;\;\;\;\;\;\;\;\;\;\;\;\;\;\;\;\;\;\;\;\;\;\;\;\;\;\;\;\;\;\;\;\;\;\;\;\;\;\;-\Bigg( {\sum\limits_{l = 1}^{{c _{m'}(a_n,a_{-n})}} \pi_{m'k}e^{-\theta_n \frac{s_{m'k}}{l}}}+{\sum\limits_{l = 1}^{{c_m}(a_n,a_{-n})} \pi_{mk}e^{-\theta_n \frac{s_{mk}}{l}}} \Bigg)\\
\;\;\;\;\;\;\;\;\;\;\;\;\;\;\;\;\;\;\;\;\;\;\;\;\;\;\;\;\;\;\;\;\;\;\;\;\;\;\;\;\;\;\;\;\;\;\;\;
=\pi_{m'k}e^{-\theta_n \frac{s_{m'k}}{c_{m'}(a_n,a_{-n})+1}}-\pi_{mk}e^{-\theta_n \frac{s_{mk}}{c_m(a_n,a_{-n})}}
\end{array}
\end{equation}

\begin{equation}
\label{eq:potential_change_v3}
v_n^{(k)}(a'_n,a_{-n})-v_n^{(k)}(a_n,a_{-n})= \phi_v^{(k)} ({a'_n},{a_{ - n}})-\phi_v^{(k)} ({a_n},{a_{ - n}}), \forall n, k, a_n, a'_n
\end{equation}

 \begin{equation}
\label{eq:multiply2}
  \Big[- \frac{1}{\theta_n} \log \big(v_n(a_n,a_{-n})\big)+\frac{1}{\theta_n} \log \big(v_n(a'_n,a_{-n})\big) \Big] \Big[v_n(a_n,a_{-n})-v_n(a'_n,a_{-n})\Big]\leq 0, \forall a_n, a'_n
 \end{equation}

  \begin{equation}
\label{eq:multiply3}
  \Big[- \frac{1}{\theta_n} \log \big(\phi_v(a_n,a_{-n})\big)+\frac{1}{\theta_n} \log \big(\phi_v(a'_n,a_{-n})\big) \Big] \Big[\phi_v(a_n,a_{-n})-\phi_v(a'_n,a_{-n})\Big] \leq 0, \forall a_n, a'_n
 \end{equation}

\rule{\linewidth}{1pt}
 \end{figure*}
Accordingly, the change in $\phi_v^{(k)} ({a_n},{a_{ - n}})$ is given by (\ref{eq:potential_change_v}), which is  shown in the top of next page. The change in the channel selection of player $n$ only affects the users in channel $m$ and $m'$. Furthermore, we have
${c _{m'}(a'_n,a_{-n})}={c _{m'}(a_n,a_{-n})+1}$ and ${c _{m}(a'_n,a_{-n})}={c _{m}(a_n,a_{-n})-1}$. Therefore,
(\ref{eq:potential_change_v}) can be further expressed as (\ref{eq:potential_change_v2}).
Combining (\ref{eq:potential_function}) and (\ref{eq:potential_change_v2}), the changes in $v_n^{(k)}(a'_n,a_{-n})$ and $\phi_v^{(k)} ({a_n},{a_{ - n}})$ are related by  (\ref{eq:potential_change_v3}). Therefore, for all $n \in \mathcal{N}$, all $a_n \in \mathcal{A}_n$ and $a'_n \in \mathcal{A}_n$, it always holds that
\begin{equation}
\label{eq:utility_equal_potential_v}
v_n(a'_n,{a_{ - n}}) - v_n({a_n},{a_{ - n}})= \phi_v(a'_n,{a_{ - n}}) - \phi_v({a_n},{a_{ - n}}).
\end{equation}

Furthermore, due to the monotony of the logarithmic function, i.e., $-\frac{\log (x)}{\theta_n}$ is monotonously decreasing with respect to $x$,
the inequalities  specified by (\ref{eq:multiply2}) and (\ref{eq:multiply3}) always hold. Now, define the potential function as follows:
\begin{equation}
\label{eq:potential_u}
 \phi_u(a_n,a_{-n})=- \frac{1}{\theta_n} \log \big(\phi_v(a_n,a_{-n})\big),
 \end{equation}
 where $\phi_v(a_n,a_{-n})$ is given by (\ref{eq:potential_function_v}). Also, according to (\ref{eq:utility_function}) and (\ref{eq:utility_v}), the utility function can be re-written as follows:
 \begin{equation}
\label{eq:utility_relation}
 u_n(a_n,a_{-n})=- \frac{1}{\theta_n} \log \big(v_n(a_n,a_{-n})\big)
 \end{equation}

Then, combining  (\ref{eq:utility_equal_potential_v}), (\ref{eq:multiply2}), (\ref{eq:multiply3}), (\ref{eq:potential_u}) and (\ref{eq:utility_relation}) yields the following inequality:
\begin{equation}
\big(u_n(a'_n,{a_{ - n}}) - u_n({a_n},{a_{ - n}})\big) \big( \phi_u(a'_n,{a_{ - n}}) - \phi_u({a_n},{a_{ - n}}) \big)\ge 0,
\end{equation}
which always holds for all $n \in \mathcal{N}$,  $a_n \in \mathcal{A}_n$ and $a'_n \in \mathcal{A}_n$. According to Definition 2, it is proved that the formulated opportunistic channel access game with QoS provisioning is an OPG with $\phi_u$ serving as an ordinal potential function. Therefore, Theorem \ref{tm:potential_game} is proved\footnote{The presented proof is for scenarios with TDMA policy. If CSMA policy is applied, (\ref{eq:utility_v}) is given by:
 \[v_n(a_n,a_{-n})= \frac{1}{c_m} \Big(\sum\nolimits_{k=1}^K {\pi_{mk}}e^{-\theta_n {s_{mk}}}\Big)+1-\frac{1}{{c_m}}\]
 Then, the same result characterized by (\ref{eq:utility_equal_potential_v}) can be obtained using similar proof lines in \cite{dynamic_selection1}. Finally, Theorem 1 can also be proved following the same methodology.}.
\end{IEEEproof}

\subsection{Multi-agent learning for achieving Nash equilibria}

Since the formulated dynamic spectrum access game is an OPG, as characterized by Theorem \ref{tm:potential_game}, it has at least one pure strategy Nash equilibrium. In the literature, there are  large number of learning algorithms for an OPG to achieve its Nash equilibria, e.g., best (better) response \cite{potential_game}, fictitious play \cite{game_book} and no-regret learning \cite{static_selection2}. However, these algorithms require the environment to be static and need to know information of other users in the learning process, which means that these algorithms can not be applied to the considered dynamic system. Based on the CODIPAS learning techniques \cite{Learning_book} (COmbined fully DIstributed PAyoff and Strategy), we propose a multi-agent learning algorithm to achieve the Nash equilibria of the formulated opportunistic channel access game in the presence of unknown, dynamic and incomplete information constraints.

 For the formulated dynamic spectrum access game with QoS provisioning, the  utility function of player $n$, characterized by (\ref{eq:utility_function}), can be re-written as:
\begin{align}
 u_n(a_n,a_{-n})=\lim\limits_{T \to \infty} -\frac{1}{T\theta_n}
\log \Big( \sum\nolimits_{i=1}^T e^{-\theta_n r_n(i) } \Big).
 \end{align}
It is seen that the utility function does not enjoy the additive property with respect to the random payoff part $r_n(i)$. On the contrary, it leads to multiplicative dynamic programming in essence \cite{Learning_book}. To cope with this problem, the following approximated  part can be obtained by  performing Taylor expansion of the logarithmic function \cite{Learning_book}:
\begin{align}
 \label{eq: approximated_utility}
 u_n(a_n,a_{-n})=\frac{1-\text{E}[e^{-\theta_n r_n(i)}]}{\theta_n}+\text{o}(r_n(i)),
  \end{align}
  where $\text{o}(r_n(i))$ is the infinitely small quantity of higher order. By omitting the logarithmic term, we define $u'_n(a_n,a_{-n})=\frac{1-\text{E} [e^{-\theta_n r_n(i)}]}{\theta_n}$, which is an approximation of $u_n(a_n,a_{-n})$. It can be proved that $u'_n(a_n,a_{-n})$ has  some important properties with $u_n(a_n,a_{-n})$. In particular, $\lim_{\theta \rightarrow 0} u'_n(a_n,a_{-n})= \text{E} [  r(i) ].$

For the expected part of $u'_n(a_n,a_{-n})$, it can be written as:
\begin{align}
  y_n(T)= \frac{1}{T+1} \sum\nolimits_{i=0}^T e^{-\theta_n r_n(i) },
 \end{align}
 which can be further re-written in the following recursive form \cite{Learning_book}:
\begin{equation}
\label{eq:recursive_equalization}
\begin{array}{l}
 y_n(T)=\big(1-\frac{1}{T+1}\big)y_n(T-1)+\frac{1}{T+1}e^{-\theta_n r_n(T) }\\
 \;\;\;\;=y_n(T-1)+\frac{1}{T+1} \big(e^{-\theta_n r_n(T) }-y_n(T-1)\big)
 \end{array}
 \end{equation}

\begin{figure}[tb!]
\rule{\linewidth}{1pt}
\emph{\textbf{Algorithm 1}:  Multi-agent learning algorithm for dynamic spectrum access with QoS provisioning}
\\
\rule{\linewidth}{1pt}
\begin{algorithmic}
\STATE \textbf{Initialization:} set the iteration index $i=0$,  the initial channel selection probability vector as $\mathbf{p}_n(0)=(\frac{1}{M},\ldots, \frac{1}{M})$, and  the initial estimation $Q_{nm}(0)=0, \forall n, m$.
Let each player $n$ randomly select a channel $a_n(0)\in \mathcal{A}_{n}$ with equal probabilities.
\STATE \textbf{Loop for $i=0,1,\ldots,$}
\\
\STATE \quad \textbf{Channel access and get random payoff}: with the channel selection profile $(a_n(i),a_{-n}(i))$, the players contend for the channels and get random payoffs $r_n(i)$, which are determined by (\ref{eq:random_rate_TDMA}) or (\ref{eq:random_rate_CSMA}).
\\
\STATE \quad \textbf{Update  estimation:}
each player updates the estimations according to the following rules:
\begin{equation}
 \label{eq:update_estimation}
 \begin{array}{l}
  Q_{nm}(i+1) = Q_{nm}(i)\\
  \;\;\;\;\;\;\;\;\;\;\;\;\;\;\;\;\;\;\;\;\;\;\;\;+\lambda_i  I(a_n(i),m) \Big(\frac{1-e^{-\theta_n r_n(i)}}{\theta_n}-Q_{nm}(i)\Big),
  \end{array}
 \end{equation}
where $\lambda_i$ is the step factor, $I(a_n(i),m)=1$ if $a_n(i)=m$ and $I(a_n(i),m)=0$ otherwise.

\STATE \quad \textbf{Update channel selection probabilities:} each player updates its channel selection probabilities using the following rule:
\begin{equation}
 \label{eq:update_probability}
  p_{nm}(i+1) = \frac{p_{nm}(i) (1+\eta_i)^{Q_{nm}(i)}} {\sum\nolimits_{m'=1}^M p_{nm'}(i) (1+\eta_i)^{Q_{nm'}(i)}}, \forall n, m
 \end{equation}
where $\eta_i$ is the learning parameter. Based on the updated mixed strategy, the players choose the channel selection $a_n(i+1)$ in the next iteration.

\STATE \textbf{End loop }
 \end{algorithmic}
 \rule{\linewidth}{1pt}
\end{figure}

\begin{figure*}[bt!]
  \center
  \includegraphics[width=5.9in]{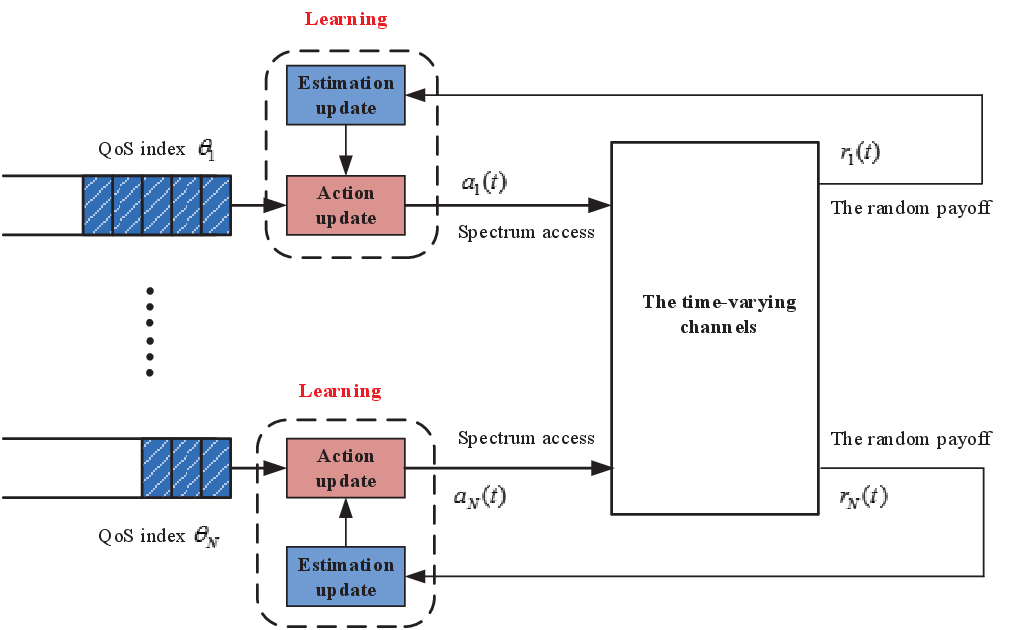}\\
  \caption{The illustrative paradigm of the multi-agent learning algorithm for dynamic spectrum access with QoS provisioning.}
  \label{fig:paradgim}
\end{figure*}

Based on the above recursive analysis, we propose a multi-agent learning algorithm, which is derived from the CODIPAS learning techniques \cite{Learning_book}. To begin with, we extend the formulated dynamic spectrum access game into a mixed strategy form. Let $\mathbf{P}(i)=(\mathbf{p}_1(i),\ldots,\mathbf{p}_n(i))$ denotes the mixed strategy profile in slot $i$, where $\mathbf{p}_n(i)=({p}_{n1}(i),\ldots,{p}_{nM}(i))$ is the probability vector of player $n$ choosing the channels. The underlying idea of the proposed multi-agent learning algorithm is that each player chooses a channel, receives a random payoff, and then updates its channel selection in the next slot. Specifically, it can be summarized as follows:
 i) in the first slot, each player chooses the channels with equal probabilities, i.e., $\mathbf{p}_n(0)=(\frac{1}{M},\ldots,\frac{1}{M}), \forall n\in{\mathcal{N}}$, ii) at the end of slot $t$, player $n$ receives random payoff $r_n(t)$ and constructs estimation $Q_{nm}$ for the aggregate reward of choosing each channel, and iii) it updates its mixed strategy based on the estimations. Formally, the illustrative paradigm of the multi-agent learning algorithm for dynamic spectrum access with QoS provisioning is shown in Fig. \ref{fig:paradgim} and the procedure is formally described in Algorithm 1.

 The properties of the proposed multi-agent learning algorithm are characterized by the following theorems. First, using the method of ordinal differential equalization (ODE) approximation, the long-term behaviors of the probability matrix sequence $\mathbf{P}(i)$  and the estimation sequence $Q(i)$ are characterized. Secondly, the stable solutions of the approximated ODE are analyzed.

 To begin with,  define $\omega_n(m,\mathbf{p}_{-n})$ as the expected value of $u'_n(a_n,a_{-n})$ when player $n$ chooses channel $m$ while all other players choose their channels according to the mixed strategies, i.e.,
 \begin{equation}
\label{eq:expect_utility}
\begin{array}{l}
 \omega_n(m,\mathbf{p}_{-n})=\text{E}_{a_{-n}}[ u'_n(a_n,a_{-n})]|_{a_n=m}\\
   \;\;\;\;=\sum\limits_{a_k,k\ne n}u'_n(a_1,\ldots,a_{n-1},m,a_{n+1},\ldots,a_{N}) \prod\limits_{k \ne n} p_{ka_k}
 \end{array}
 \end{equation}

\newtheorem{proposition}{Proposition}
\begin{proposition}
\label{tm:ODE}
With sufficiently small $\lambda_i$ and $\eta_i$, the  channel selection probability matrix sequence $p_{nm}(i)$ can be approximately characterized by the following ODE:
\begin{align}
\label{eq:ODE}
\begin{array}{l}
  \frac{dp_{nm}(t)}{dt}=p_{nm}(t) \Big[\omega_n(m,\mathbf{p}_{-n})\\
   \;\;\;\;\;\;\;\;\;\;\;\;\;\;\;\;\;\;\;\;\;\;\;\;\;\;\;\;-\sum\nolimits_{m'=1}^M p_{nm'}(t)\omega_n(m',\mathbf{p}_{-n})\Big]
 \end{array}
\end{align}
\end{proposition}

\begin{IEEEproof}
The following proof follows the lines for  proof in \cite{Learning_book}, which are mainly based the theory of stochastic approximation.

First, the expected changes of the estimation $Q_{nm}(i)$ in one slot is as follows:
\begin{equation}
 \label{eq:ODE_1}
 \begin{array}{l}
 \text{E}\Big( \frac{Q_{nm}(i+1)-Q_{nm}(i)}{\lambda_i}|\mathbf{p}_n(i) \Big)\\
 \;\;=p_{nm}(i) \Big(\frac{1-\text{E}[e^{-\theta_n r_n(i)}]}{\theta_n}-Q_{nm}(i)\Big).
 \end{array}
 \end{equation}
 If the step factor $\lambda_i$ is sufficiently small, the discrete time process (\ref{eq:ODE_1}) can be approximated by the following differential equalization:
\begin{equation}
 \label{eq:ODE_2}
 \begin{array}{l}
 \frac{dQ_{nm}(t)}{dt}=p_{nm}(i) \Big(\frac{1-\text{E}[e^{-\theta_n r_n(t)}]}{\theta_n}-Q_{nm}(t)\Big).
 \end{array}
 \end{equation}

Second, the changes of the channel selection probability in one slot is as follows:
\begin{equation}
 \label{eq:ODE_3}
 \begin{array}{l}
 \frac{p_{nm}(i+1)-p_{nm}(i)}{\eta_i}\\
 =\frac{1}{\eta_i} \Big[\frac{p_{nm}(i) (1+\eta_i)^{Q_{nm}(i)}} {\sum\nolimits_{m'=1}^M p_{nm'}(i) (1+\eta_i)^{Q_{nm'}(i)}}-p_{nm}(i) \Big]\\
 = \frac{p_{nm}(i) } {\sum\nolimits_{m'=1}^M p_{nm'}(i) (1+\eta_i)^{Q_{nm'}(i)}}\\
 \;\;\;\;\times\Big[ \frac{(1+\eta_i)^{Q_{nm}(i)}-1}{\eta_i}-\sum\limits_{m'=1}^M p_{nm'}(i) \frac{(1+\eta_i)^{Q_{nm'}(i)}-1}{\eta_i}\Big].
 \end{array}
 \end{equation}
Using the fact that $\frac{(1+\eta_i)^{x}-1}{\eta_i} \to x$ as $\eta_i \to 0$,  and taking the conditional expectation, the discrete time process (\ref{eq:ODE_2}) can be approximated by the following differential ordinal equalization:
\begin{equation}
 \label{eq:ODE_4}
 \begin{array}{l}
 \frac{dp_{nm}(t)}{dt}=p_{nm}(t) \Big( \text{E}[Q_{nm}(t)] -\sum\nolimits_{m'=1}^M p_{nm'}(t)  \text{E}[Q_{nm'}(t)]\Big).
 \end{array}
 \end{equation}

Furthermore, according to the asymptotic convergence of the estimation updating process \cite{Learning_book}, we have $\text{E}[Q_{nm}(t)] \to \omega_n(m,\mathbf{p}_{-n})$ for (\ref{eq:ODE_4}). Therefore, Theorem \ref{tm:ODE} is proved.
\end{IEEEproof}

For the proposed multi-agent algorithm, the stable solutions of (\ref{eq:ODE}) and the Nash equilibria of the formulated channel access game with approximated utility function $u'_n(a_n,a_{-n})$ are related by the following proposition \cite{Learning_book,learning_automata}.

\begin{proposition}
The following statements are true for the proposed multi-agent algorithm:
\begin{enumerate}
  \item All the stable stationary points of the ODE are Nash equilibria.
  \item All Nash equilibria are the stationary points of the ODE.
\end{enumerate}
\end{proposition}

\begin{theorem}
\label{tm:convergence}
With sufficiently small $\lambda_i$ and $\eta_i$, the  proposed multi-agent algorithm asymptotically converges to Nash equilibria of the formulated dynamic spectrum access game with approximated utility function $u'_n(a_n,a_{-n})$.
\end{theorem}
\begin{IEEEproof}
The proof follows the lines for proof in \cite{dynamic_selection1, Learning_book}.
It is seen that $u'_n(a_n,a_{-n})=\frac{1-v_n(a_n,a_{-n})}{\theta_n}$, where $v_n(a_n,a_{-n})$ is defined in (\ref{eq:utility_v}). Therefore, there also exists an ordinal potential function for $u'_n(a_n,a_{-n})$. Specially, the potential function for $u'_n(a_n,a_{-n})$ is expressed as:
\begin{equation}
 \label{eq:potential_function_u'}
 \phi_{u'}(a_n,a_{-n})=\frac{1-\phi_v(a_n,a_{-n})}{\theta_n},
 \end{equation}
where $\phi_v(a_n,a_{-n}$ is characterized by (\ref{eq:potential_function_v}).

We define the expected value of the potential function over mixed strategy profile $\mathbf{P}$ as
$\Phi_{u'}( \mathbf{P})$ and the expected value of the potential function when player $n$ chooses a pure strategy $m$ while all other active players employ mixed strategies $\mathbf{p}_{-n}$ as $\Phi_{u'}( m,\mathbf{p}_{-n})$.
Since $\Phi_{u'}( \mathbf{P})=\sum\nolimits_{m} p_{nm}\Phi_{u'}(m,\mathbf{p}_{-n})$,
the variation of $\Phi_{u'}(\mathbf{P})$ can be expressed as follows:
\begin{equation}
\frac{\partial \Phi_{u'}( \mathbf{P})} {\partial p_{nm}}=\Phi_{u'}(m,\mathbf{p}_{-n})
\end{equation}

 We can re-write the ODE specified by (\ref{eq:ODE}) as follows:
\begin{align}
\label{eq:ODE2}
\begin{array}{l}
  \frac{dp_{nm}(t)}{dt}=p_{nm}(t) \Big[ \sum \nolimits_{m'=1}^M p_{nm'} \omega_n(m,\mathbf{p}_{-n}) \\ \;\;\;\;\;\;\;\;\;\;\;\;\;\;\;\;\;\;\;\;\;\;\;\;\;\;\;\;\;-\sum\nolimits_{m'=1}^M p_{nm'}(t)\omega_n(m',\mathbf{p}_{-n})\Big]
  \end{array}
\end{align}

\begin{figure*}[tb!]

\begin{align}
 \begin{array}{l}
 \label{eq:derivation}
\frac{d\Phi_{u'}( \mathbf{P})}{dt}=\sum_{n,m}\frac{\partial \Phi_{u'}( \mathbf{P})} {\partial p_{nm}} \frac{dp_{nm}}{dt}\\
=\sum_{n,m} \Phi_{u'}(m,\mathbf{p}_{-n}) p_{nm}(t) \Big[ \sum \nolimits_{m'=1}^M p_{nm'} \omega(m,\mathbf{p}_{-n}) -\sum\nolimits_{m'=1}^M p_{nm'}(t)\omega(m',\mathbf{p}_{-n})\Big]\\
=\sum_{n,m} \Phi_{u'}(m,\mathbf{p}_{-n}) p_{nm}(t) \sum \nolimits_{m'=1}^M p_{nm'} \Big[ \omega(m,\mathbf{p}_{-n}) -\omega(m',\mathbf{p}_{-n})\Big]\\
=\sum_{n,m,m'} p_{nm}(t) p_{nm'} \Phi_{u'}(m,\mathbf{p}_{-n})\Big[ \omega(m,\mathbf{p}_{-n}) -\omega(m',\mathbf{p}_{-n})\Big]\\
=\frac{1}{2}\sum_{n,m,m'} p_{nm}(t) p_{nm'} \big[\Phi_{u'}(m,\mathbf{p}_{-n})-\Phi_{u'}(m',\mathbf{p}_{-n})\big] \Big[ \omega(m,\mathbf{p}_{-n})- \omega(m',\mathbf{p}_{-n})\Big]\\
\end{array}
\end{align}

\rule{\linewidth}{1pt}
 \end{figure*}

The derivation of $\Phi_{u'}( \mathbf{P})$ is given by  (\ref{eq:derivation}), which is shown in the top of next page. According to the properties of EPG and OPG, we have:
\begin{align}
\begin{array}{l}
 \big[\Phi_{u'}(m,\mathbf{p}_{-n})-\Phi_{u'}(m',\mathbf{p}_{-n})\big] \\
 \;\;\;\;\;\times \Big[ \omega_n(m,\mathbf{p}_{-n}) -\omega_n(m',\mathbf{p}_{-n})\Big]>0
 \end{array}
\end{align}

Therefore, we have $\frac{d\Phi_{u'}( \mathbf{P})}{dt}\ge0$, which implies that $\Phi_{u'}(\mathbf{P})$ increases as the algorithm iterates. Furthermore, since $\Phi_{u'}(\mathbf{P})$ is upper-bounded, it will eventually converge to some maximum points, as $\frac{d\Phi_{u'}(\mathbf{P})}{dt}=0$. Finally, we have the following relationships:
\begin{align}
\begin{array}{l}
\frac{d\Phi_{u'}(\mathbf{P})}{dt}=0\\
\Rightarrow  \omega_n(m,\mathbf{p}_{-n}) -\omega_n(m',\mathbf{p}_{-n})=0, \forall n,m,m'\\
\Rightarrow \frac{dp_{nm}}{dt}=0, \forall n,m\\
 \Rightarrow\frac{d\mathbf{P}}{dt}=0\\
\end{array}
\end{align}

The last equation shows that $\mathbf{P}$ eventually converges to the stationary point of (\ref{eq:ODE}). Therefore, according to Proposition 2, it is proved that the proposed multi-agent learning algorithm converges to Nash equilibria of the formulated opportunistic channel access game with approximated utility function $u'_n(a_n,a_{-n})$., which proves Theorem \ref{tm:convergence}.
\end{IEEEproof}

\textbf{Remark 1.}
 It is noted the proposed algorithm is distributed and uncoupled, i.e., each player makes the decisions autonomously and does not to know information about other players. However, it should be pointed out that the estimation  of Q-value, which is originally derived from the recursive equalization  (\ref{eq:recursive_equalization}), is not equal to the expectation of effective capacity. Actually, the Q-value is used to represent the winningness for a user choosing a particular channel. More specifically, each user chooses a channel according to the Q-values and then updates them based on the outcome of last channel selections. In general, a user prefers to choose channels with high Q-values. As the users randomly change their channel selections, the Q-values are also updated randomly in time. Thus, the estimation is generally not equal to the actual expectation.

As no prior information is available in the initial stage, the users chooses the channels with equal probabilities, i.e., the Q-values for all the channels are set to the same. If the initial choices are different, the algorithm still converges to a stable solution. However, different initial parameters may result in different stable solutions. The reasons are as follows: the instantaneous channel rates, the channel selection profiles and the user contentions are random, which leads to the random payoff after each play. Then, the converging channel selection is also not deterministic.

 \textbf{Remark 2.}
The choice of $\lambda$ is to balance the tradeoff between exploration and exploitation. In practice, the value of $\lambda$ decreases as the algorithm iterates. Specifically, in the beginning state, the users want to explore all channels and hence the Q-values on each channel are updated significantly. However, as the algorithm iterates, the users want to exploit the best channel and the Q-values are updated trivially. In practice, we can use $\lambda =1/t$, where $t$ is the iteration index. Also,  the value choice of $\eta$ balances the tradeoff between performance and convergence speed. For larger value of $\eta$, it converges rapidly but it may converge to local optimal solutions. For smaller values, it has more opportunities to find global optimal solutions but it may take more times. Thus, the learning parameters should be application-dependent \cite{dynamic_selection1}.

 \textbf{Remark 3.} Although the above  convergence analysis is for the game with the approximated utility function $u'_n$, the convergence for the original game can be expected. The reason is that the approximated utility function is close to the original utility function. In particular, its convergence will be verified by simulation results in the next section.

\section{Simulation Results and Discussion}
We use the finite state channel model  to characterize the time-varying transmission rates of the channels. Specifically, with the help of adaptive modulation and coding (ACM), the channel transmission rate is classified into several states according to the received instantaneous signal-to-noise-ratio (SNR). The state classification is jointly determined by the average received SNR $\gamma$ and the target packet error rate $p_e$. The HIPERLAN/2 standard \cite{HIPERLAN_2} is applied in the simulation study, in which the channel  rate set is given by $\{0,1,2,3,6\}$. Here, the rate is defined as the transmitted packets  in a slot. To make it more general,
we consider Rayleigh fading and set different average SNR for the channels\footnote{It is noted that such a configuration is just for the purpose of illustration. The proposed multi-agent learning approach can applied to other scenarios.}. Using the method proposed in \cite{finite_rate_model}, the state probabilities can be obtained for a given average SNR and a certain packet error rate. Taking $\gamma=5$ dB and $p_e=10^{-3}$ as an example, the state probabilities are given by $\pi=\{0.3376, 0.2348, 0.2517, 0.1757, 0.002\}$.
 Furthermore, the learning parameters are set to $\lambda_i=\frac{1}{t}$ and $\eta_i=0.1$ unless otherwise specified.
 The CSMA policy is applied in the simulation study\footnote{Due to the limited space, we only present simulation results for CSMA policy here. Simulation results for TDMA policy can be found in \cite{ICC17_Xu}, which admits similar tendencies as expected.}. We first present the convergence behaviors of the proposed  multi-agent learning algorithm, and then investigate the effective capacity performance.

\subsection{Convergence behavior}
In this subsection, we study the convergence behavior of the proposed multi-agent learning approach. Specifically, there are eight users and
five channels with average received SNR being 5dB, 6dB, 7dB, 8dB and 9dB respectively. For convenience of presentation, the QoS indices of all the users are set to $\theta=10^{-2}$.

 \begin{figure}[bt!]
  \center
  \includegraphics[width=3.0in]{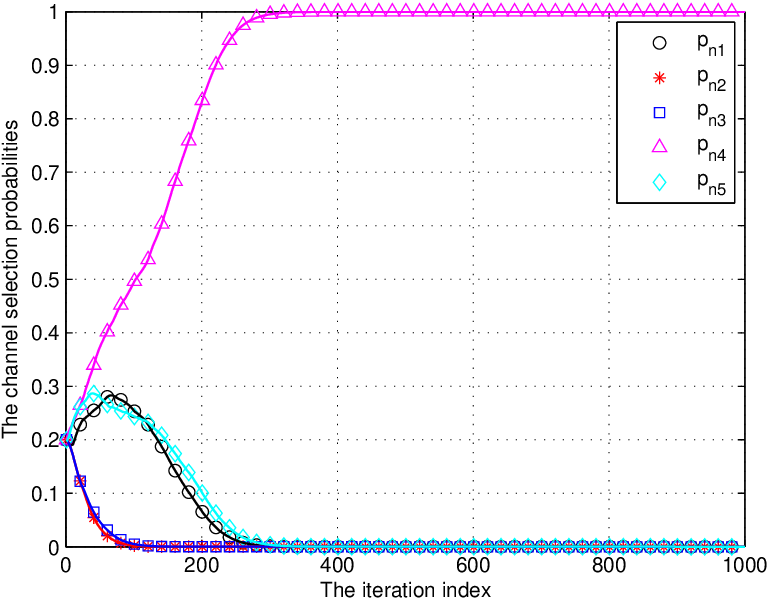}\\
  \caption{The evolution of channel selection probabilities of an arbitrarily chosen user (the number of users is $N=5$ and the QoS indices of the users are set to $\theta=10^{-2}$).}
  \label{fig:convergence_pro}
\end{figure}

\begin{figure*}[!t]
\begin{minipage}[t]{0.33\linewidth}
\centering
\includegraphics[width=2.0in]{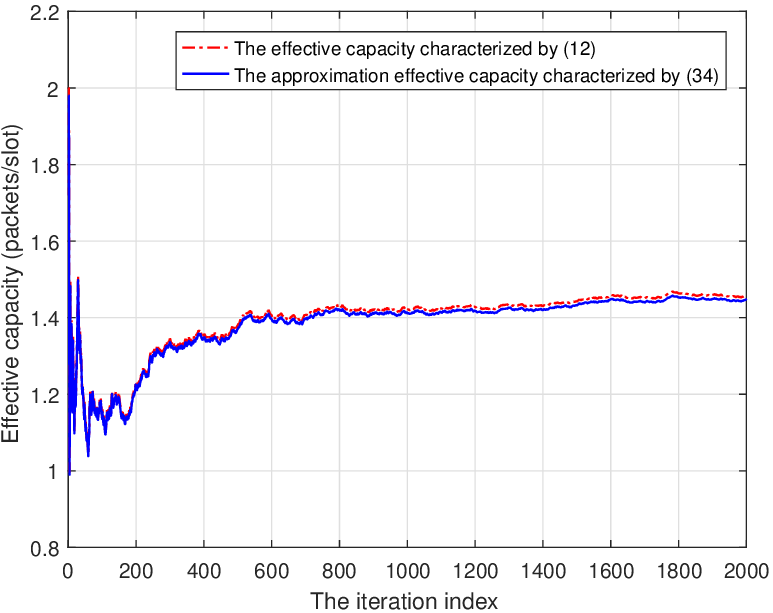}
\caption{Small QoS index ($\theta=10^{-2}$).}
\label{fig:convergence_QoS1}
\end{minipage}
\begin{minipage}[t]{0.33\linewidth}
\centering
\includegraphics[width=2.0in]{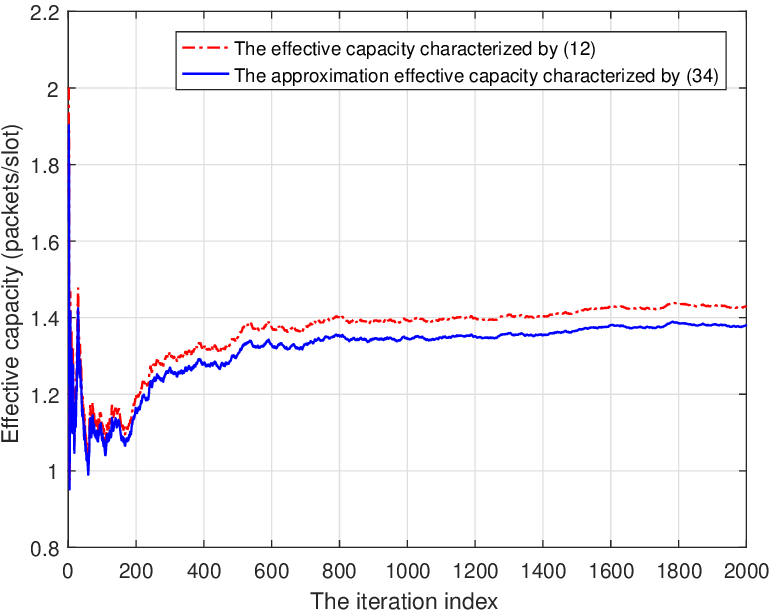}
\caption{ Moderate QoS index ($\theta=5 \times 10^{-2}$).}
\label{fig:convergence_QoS2}
\end{minipage}
\begin{minipage}[t]{0.3\linewidth}
\centering
\includegraphics[width=2.0in]{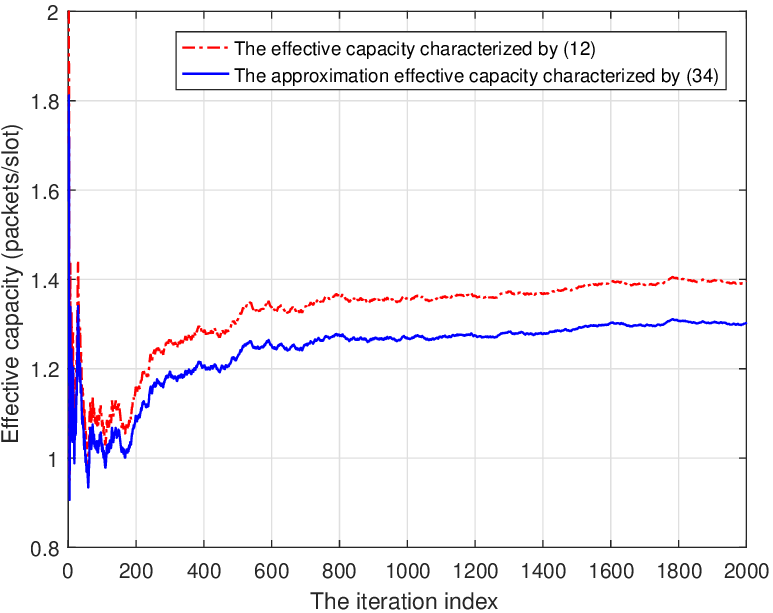}
\caption{Large QoS index ($\theta=10^{-1}$).}
\label{fig:convergence_QoS3}
\end{minipage}
\end{figure*}

For an arbitrarily chosen user, the evolution of channel selection probabilities are shown in Fig. \ref{fig:convergence_pro}. It is noted that the selection properties converge to a pure strategy (\{0,0,0,1,0\}) in about 400 iterations. These results validate the convergence of the proposed multi-agent learning algorithm with  dynamic and incomplete information. In addition, for an arbitrarily choosing user, the evolution of effective capacity, as characterized by (\ref{eq:utility_function}), and the approximation effective capacity, as characterized by (\ref{eq: approximated_utility}), are shown in Fig. \ref{fig:convergence_QoS1}-\ref{fig:convergence_QoS3}. It is noted from the figures that the effective capacity also converges in about 500 iterations. The interesting results are: i) for small QoS index, e.g., $\theta=10^{-2}$, the effective capacity is almost the same with the proposed approximation effective capacity, ii) for moderate QoS index, e.g.,  $\theta=5\times 10^{-2}$, there is a slight performance gap (less than 0.05), iii) for large  QoS index, e.g., although the  performance gap increases, it is still acceptable (about 0.1). These results validate not only the convergence of the proposed learning algorithm but also the effectiveness of the proposed approximation formulation.

\begin{figure}[bt!]
  \center
  \includegraphics[width=3.0in]{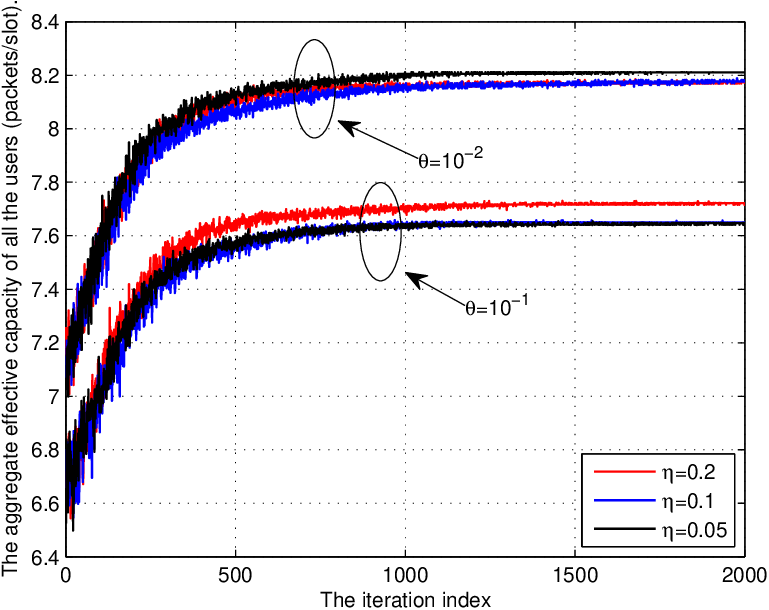}\\
  \caption{The convergence behaviors versus different learning parameter $\eta$ for different QoS indices (the number of users is $N=8$).}
  \label{fig:convergence_eta}
\end{figure}

We study the convergence behavior versus the learning parameter $\eta$ for different QoS indices and the comparison results for different parameters are shown in Fig. \ref{fig:convergence_eta}. These results are obtained by performing 200 independent trials and then taking the expectation. It is noted from the figure that the convergence behaviors are different for different QoS indices. In particular, for relatively small QoS indices, e.g., $\theta=10^{-2}$, the final achievable performance increases as the learning parameter $\eta$ decreases. On the contrary, for relatively large QoS indices, e.g., $\theta=10^{-1}$, the trend is opposite. Also, it is noted although it takes about 2000 iterations for the proposed multi-agent learning algorithm to converge, it achieves satisfactory performance rapidly (e.g., it achieves 90\% performance in about 500 iterations). Thus, the choice of the algorithm iteration is application-dependent.

\begin{figure}[bt!]
  \center
  \includegraphics[width=3.5in]{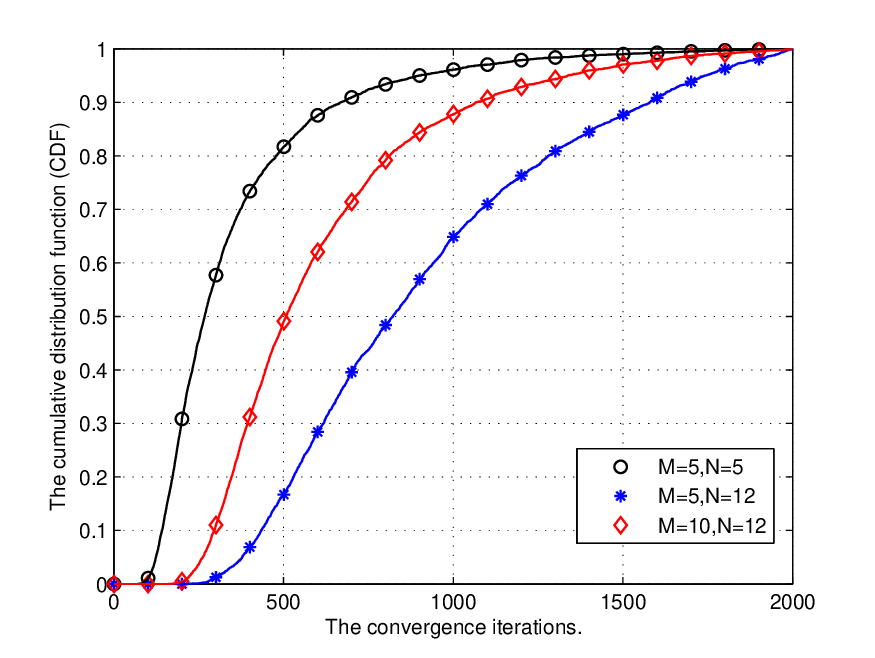}\\
  \caption{The communicative distribution function (CDF) of convergence iterations for different number of users ($N$) and number of channels ($M$).}
  \label{fig:convergence_cdf}
\end{figure}
As the convergence iterations is random, we study its cumulative distribution function (CDF) in Fig. \ref{fig:convergence_cdf}. It is shown that for a given number of channels, e.g., $M=5$, increasing the number of users (for example from $N=5$ to $N=12$) decreases the convergence speed. Also, for a given number of users, e.g., $N=12$, increasing the number of channels (for example from $M=5$ to $M=10$) accelerates the convergence speed. The reasons is as follows: when the number of users increases, the spectrum becomes crowded and hence  it needs more time to converge.

\subsection{Throughput performance}
In this subsection, we evaluate the throughput performance of the proposed multi-agent learning algorithm. We study the achievable effective capacities of the users with different QoS indices. Furthermore, we compare the proposed multi-agent learning algorithm with the random selection approach. Under the dynamic and incomplete information, random selection is an instinctive approach. For convenience of simulating, the QoS indices of all the users are set to the same, otherwise specified.

\begin{figure}[bt!]
  \center
  \includegraphics[width=3.5in]{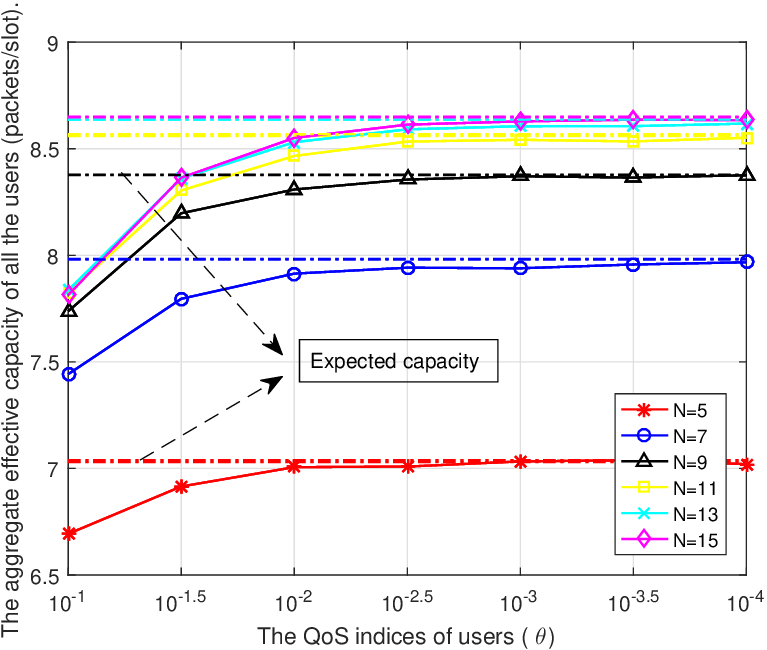}\\
    \caption{The achievable aggregate effective capacity of the uses for different statistical QoS indices.}
  \label{fig:comparison_QoS}
\end{figure}
To begin with, the achievable effective capacities of the users with different QoS indices are shown in Fig. \ref{fig:comparison_QoS}. There are also five channels with the average received SNR being 5dB, 6dB, 7dB, 8dB and 9dB respectively. The results are obtained by taking 5000 independent trials and then taking expectation.
It is noted that for a given QoS index, e.g., $\theta=10^{-2}$, increasing the number of the users leads to significant increases in the aggregate effective capacity when the number of users is small, e.g., $N\le11$. However, it is also shown that the increase in the aggregate effective capacity becomes trivial when the number of users is large, e.g., $N >11$. The reason is that the access opportunities are abundant when the number of the users is small while they are saturated when the number of users is large. Also, for a given number of users, e.g., $N=7$, the achievable aggregate increases as the QoS indices decrease. In particular, as the QoS indices become sufficiently small, e.g., $\theta<10^{-3}$, the achievable effective capacity moderates. The reasons are as follows: 1) smaller value of QoS index implies loose QoS requirements in the packet violating probability and hence leads to larger effective capacity, and 2) when the QoS index approaches zero, say,
when it becomes sufficiently small, the effective capacity degrades to the expected capacity. It is noted that the presented results in this figure comply with the properties of the effective capacity, which were analyzed in Section III.B.

\begin{figure}[bt!]
  \center
  \includegraphics[width=3.0in]{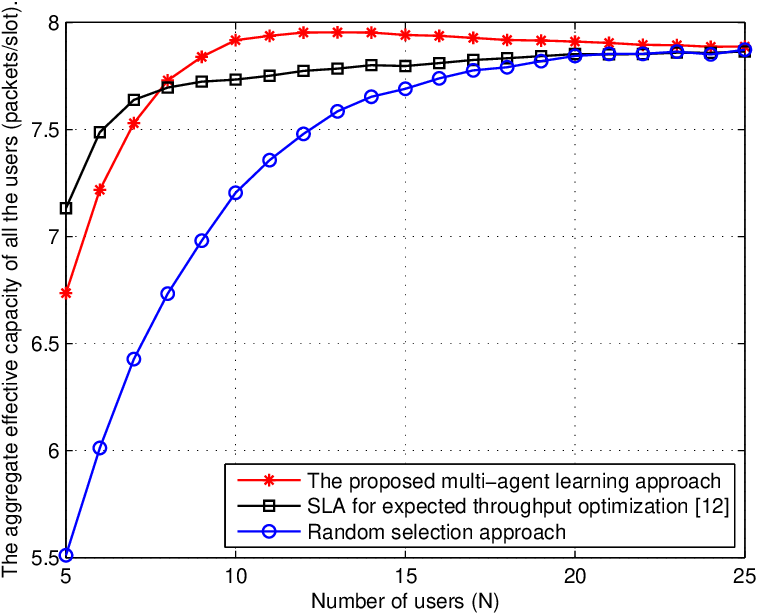}\\
    \caption{The comparison results between the proposed multi-agent learning approach and the SLA approach for expected throughput optimization.}
  \label{fig:comparison_SLA}
\end{figure}

Secondly, in order to validate the proposed learning approach for effective capacity optimization, we compare it with
an existing stochastic learning automata algorithm (SLA), which is an efficient solution for expected throughput optimization in dynamic  and unknown environment \cite{dynamic_selection1}. Specifically, the SLA algorithm is implemented for maximizing the expected throughput explicitly rather than maximizing the effective capacity, and then the achievable effective capacity is calculated over the converging channel selection profile.
There are also five channels with the average received SNR being 5dB, 6dB, 7dB, 8dB and 9dB respectively, and
the QoS indices of the users are randomly chosen from the following set $A=[0.2\times10^{-1}, 0.5\times10^{-1},10^{-1}, 2\times10^{-1}, 5\times10^{-1}, 0.2\times10^{-2},0.5\times10^{-2},10^{-2},10^{-3}]$ and the learning step size of SLA is set to $b=0.08$. The comparison results are shown in Fig. \ref{fig:comparison_SLA}. It is noted from the figure that the performance of the proposed learning algorithm is better than the SLA algorithm when $N>8$, which follows the fact that the SLA algorithm is for expected throughput optimization and is not for effective capacity optimization. However, when the number of users is small, i.e., $N<8$, the SLA approach performs better. The reasons can be analyzed as follows: (i) the competition among users is slight in this scenario, and (ii) the SLA approach converges to more efficient channel selection profiles in this scenario.
The presented results again validate the effectiveness of the proposed multi-agent learning approach for effective capacity optimization.

We also compare the proposed learning algorithm with the random selection approach in Fig. \ref{fig:comparison_SLA}. It is noted from Fig. that the achievable performance of the approaches increase rapidly as  $N$ increases when the number of users is small, e.g., $N<15$, while it becomes moderate when the number of users is large, e.g., $N>20$.
The reasons are: 1) when the multi-agent learning approach finally converges to a pure strategy, the users are spread over the channels. On the contrary, the users are in disorder with the random selection approach, which means that some channels may be crowded while some others may  be not occupied by any user. 2) the access opportunities are abundant when the number of users is small, which means that adding a user to the system leads to relatively significant performance improvement. On the contrary, the access opportunities are saturated when the number of users is large, which means that the performance improvement becomes small. 3) when the number of users becomes sufficiently larger, the users are asymptotically uniformly  spread over the channels. Thus, the performance gap between the two approaches is trivial.

\begin{figure}[bt!]
  \center
  \includegraphics[width=3.0in]{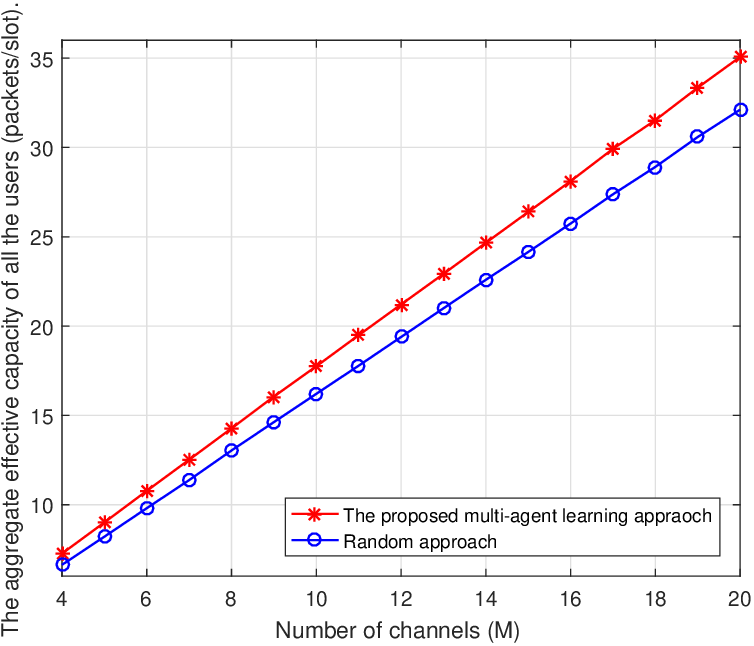}\\
    \caption{The comparison results between the proposed multi-agent learning approach and random approach with  fixed $N/M=2$ (the QoS indices are set as $\theta=10^{-2}$).}
  \label{fig:comparison_loose}
\end{figure}

Thirdly, considering the tendency in future wireless networks where the networks are dense and more resources are available \cite{Xu_magazine} (e.g. use of high frequencies), we evaluate the proposed learning algorithm for denser networks with fixed ratio of number of users and channels $N/M=2$ in Fig. \ref{fig:comparison_loose}. The QoS indices of the users are set to $\theta=10^{-2}$, and the average received SNR of the channels are randomly selected from $[5\text{dB},6\text{dB},7\text{dB},8\text{dB},9\text{dB}, 10 \text{dB}]$ in each trial. The results are obtained by taking 5000 independent trials and then taking expectation. It it noted from the figure that both approaches increase linearly as the number of channels increases. Also, it is noted from the figure that the proposed multi-agent learning algorithm significantly outperforms the random selection approach while the performance gap increases as the number of users increases. These results validate the effectiveness of the proposed learning algorithm in future wireless networks.

%
%


\section{Conclusion}
In this article, we investigated the problem of dynamic spectrum access in time-varying environment. To capture the expectation and fluctuation in dynamic environment, we considered effective capacity, which takes into account not only the expectation but also other-order moments, to characterize the statistical QoS constraints in packet delay. We formulated the interactions among the users in the dynamic environment as a non-cooperative game and proved it is an ordinal potential game which has at least one pure strategy Nash equilibrium. Based on an approximated utility function, we proposed a multi-agent learning algorithm which is proved to achieve stable solutions with  dynamic and incomplete information constraints. The convergence of the proposed learning algorithm is verified by simulation.  In future, we plan to establish a general distributed optimization framework which considers the expectation and  other higher-order moments.

Due to the fact that the considered dynamic spectrum access network is fully distributed and autonomous, NE solution is desirable in this work. However, when information exchange is available, some more efficient solutions beyond NE, e.g., the before-mentioned Nash bargaining and coalitional games, should be developed. In future work, we also plan to develop solutions beyond NE for spectrum management in dense dynamic and heterogeneous networks, in which there is a controller in charge for the users and information exchange is feasible.

\ifCLASSOPTIONcaptionsoff
  \newpage
\fi

\bibliographystyle{IEEEtran}
\bibliography{IEEEabrv,reference}

\begin{thebibliography}{10}
\providecommand{\url}[1]{#1}
\csname url@samestyle\endcsname
\providecommand{\newblock}{\relax}
\providecommand{\bibinfo}[2]{#2}
\providecommand{\BIBentrySTDinterwordspacing}{\spaceskip=0pt\relax}
\providecommand{\BIBentryALTinterwordstretchfactor}{4}
\providecommand{\BIBentryALTinterwordspacing}{\spaceskip=\fontdimen2\font plus
\BIBentryALTinterwordstretchfactor\fontdimen3\font minus
  \fontdimen4\font\relax}
\providecommand{\BIBforeignlanguage}[2]{{%
\expandafter\ifx\csname l@#1\endcsname\relax
\typeout{** WARNING: IEEEtran.bst: No hyphenation pattern has been}%
\typeout{** loaded for the language `#1'. Using the pattern for}%
\typeout{** the default language instead.}%
\else
\language=\csname l@#1\endcsname
\fi
#2}}
\providecommand{\BIBdecl}{\relax}
\BIBdecl




\bibitem{attention1}
J.~Mitola, and G.~Q.~Maguire, ``Cognitive radio: making software radios more personal,''
 \emph{IEEE Personal Communications}, vol.~6, no.~4, pp. 13 --18, Aug. 1999.


\bibitem{attention3}
S.~Haykin, ``Cognitive radio: brain-empowered wireless communications,''
  \emph{{IEEE} J. Sel. Areas Commun.}, vol.~23, no.~2, pp. 201 -- 220, 2005.


\bibitem{Decision_Theoretic_CR}
Y.~Xu, A.~Anpalagan, Q.~Wu, et al., ``Decision-theoretic distributed channel selection for
opportunistic spectrum access: Strategies, challenges and solutions,"  \emph{IEEE Communications Surveys and Tutorials},
vol. 15, no. 4, pp. 1689-1713, Fourth Quarter, 2013.

\bibitem{Spectrum_Decision_CR}
M.~Masonta, M.~Mzyece, N.~Ntlatlapa, ``Spectrum decision in cognitive radio networks:
A survey," \emph{IEEE Communications Surveys and Tutorials}, vol.~15, no.~3, pp.~1088--1107, 2013.





\bibitem{DSA_CR}
Ian.~F.~Akyildiz, W.~Lee, M.~Vuran, et al., ``NeXt generation/dynamic spectrum access/cognitive radio wireless networks:
 a survey," \emph{Computer Networks}, vol.~50, no.~13, pp.~2127-2159, 2006.
 \bibitem{static_selection1}

Y.~Xu, J.~Wang, Q.~Wu, et al, ``Opportunistic spectrum access in cognitive radio
networks: Global optimization using local interaction games,''
\emph{IEEE J. Sel. Signal Process}, vol.~6, no.~2, pp.~180--194, 2012.

\bibitem{static_selection2}
N.~Nie and C.~Comaniciu, ``Adaptive channel allocation spectrum etiquette for
  cognitive radio networks,'' \emph{Mobile Networks \& Applications}, vol.~11,
  no.~6, pp. 779--797, 2006.

\bibitem{static_selection3}
M.~Felegyhazi, M.~Cagalj, and J.~P.~Hubaux, ``Efficient MAC in cognitive radio
  systems: A game-theoretic approach,'' \emph{{IEEE} Trans. Wireless Commun.},
  vol.~8, no.~4, pp. 1984 --1995, 2009.

\bibitem{static_selection4}
M.~Maskery, V.~Krishnamurthy, and Q.~Zhao, ``Decentralized dynamic spectrum
  access for cognitive radios: Cooperative design of a non-cooperative game,''
  \emph{{IEEE} Trans. Commun.}, vol.~57, no.~2, pp. 459 --469, 2009.

  \bibitem{static_selection5}
Y.~Xu, Q.~Wu, J.~Wang, et al, ``Opportunistic spectrum access using partially overlapping channels:
Graphical game and uncoupled learning", \emph{IEEE Trans. on Commun.,} vol. 61, no. 9, pp. 3906-2918, 2013.

\bibitem{static_selection6}
Y.~Xu, Q.~Wu, L.~Shen, et al, ``Opportunistic spectrum access with spatial reuse: Graphical game and uncoupled
learning solutions," \emph{IEEE Trans. on Wireless Commun.,} vol.~12, no.~10, pp.4814--4826, 2013.

  \bibitem{dynamic_selection1}
Y.~Xu,  J.~Wang, and Q.~Wu, et al, ``Opportunistic spectrum access in
unknown dynamic environment: A game-theoretic stochastic learning solution",
\emph{{IEEE} Trans. Wireless Commun.}, vol.~11, no.~4, pp. 1380 --1391, 2012.

\bibitem{dynamic_selection2}
Q.~Zhao, L.~Tong, A.~Swami, and Y.~Chen, ``Decentralized cognitive MAC for opportunistic spectrum access in ad hoc networks: A POMDP framework,"
\emph{{IEEE} J. Sel. Areas Commun.}, vol.~25, no.~3, pp.~589-600, 2007

\bibitem{dynamic_selection3}
Q.~Wu, Y.~Xu, J.~Wang, et al, ``Distributed channel selection in time-varying radio environment: Interference
 mitigation game with uncoupled stochastic learning," \emph{IEEE Trans. on Veh. Technol.}, vol.~62, no.~9, pp.~4524 - 4538, 2013.


\bibitem{MAB1}
K.~Liu and Q.~Zhao, ``Distributed learning in multi-armed bandit with multiple players,"
\emph{{IEEE} Trans. Signal Process.,} vol.~58, no.~11, pp.~5667-5681, 2010.

\bibitem{effective_capacity1}
A.~Balasubramanian and S.~Miller, ``The effective capacity of a time division downlink scheduling system,"
\emph{IEEE Trans. Commun.,} vol.~58, no.~1, pp.~73-78, 2010.

\bibitem{Learning_book}
H.~Tembine, \emph{Distributed strategic learning for wireless engineers}, CRC Press, 2012.


\bibitem{potential_game}
D.~Monderer. and L.~S. Shapley, ``Potential games,'' \emph{Games and Economic
  Behavior}, vol.~14, pp. 124--143, 1996.

\bibitem{game_book}
R.~Myerson, \emph{Game Theory: Analysis of Conflict}. Cambridge, MA:
Harvard Univ. Press, 1991.



  \bibitem{effective_capacity3}
  L.~Musavian, S.~A\"{i}ssa and S.~Lambotharan, ``Effective capacity for interference and delay constrained
   cognitive radio relay channels,"  \emph{{IEEE} Trans. Wireless Commun.}, vol.~9, no.~5, pp.~1698-1707, 2010.

    \bibitem{effective_capacity4}
   L.~Musavian and S.~A\"{i}ssa, ''Effective capacity of delay-constrained cognitive radio
    in Nakagami fading channels,"   \emph{{IEEE} Trans. Wireless Commun.}, vol.~9, no.~3, pp.~1054-1062, 2010.

   \bibitem{effective_capacity5}
   S.~ Akin and M.~Gursoy,  ''Effective capacity analysis of cognitive radio channels
   for quality of service provisioning," \emph{{IEEE} Trans. Wireless Commun.}, vol.~9, no.~11, pp.~3354-3364, 2010.


   \bibitem{effective_capacity6}
    H.~Su and  X.~Zhang, ``Cross-layer based opportunistic MAC protocols for QoS provisionings
    over cognitive radio wireless networks," \emph{IEEE J. Sel. Areas Commun., } vol.~26, no.~1, pp.~118-129, 2008.

%


\bibitem{static_selection7}
H.~Li and Z.~Han, ``Competitive spectrum access in cognitive radio networks:
  Graphical game and learning,'' in \emph{Proc. IEEE WCNC}, pp. 1--6, 2010.


\bibitem{static_selection8}
Y.~Xu, Q.~Wu, J.~Wang, et al., ``Distributed channel selection in CRAHNs with heterogeneous
spectrum opportunities: A local congestion game approach,''
\emph{IEICE Trans. Commun.}, vol.~E95-B, no.~3, pp.~991--994, 2012.



 \bibitem{static_selection10}
J.~Wang, Y.~Xu, A.~Anpalagan, et al, ``Optimal distributed interference avoidance: Potential game and learning,"
\emph{Transactions on Emerging Telecommunications Technologies}, vol.~23, no.~4, pp.~317-326, 2012.

 \bibitem{Sastry94}
P.~Sastry, V.~Phansalkar and M.~Thathachar, ``Decentralized learning of nash
  equilibria in multi-person stochastic games with incomplete information,''
  \emph{{IEEE} Trans. Syst., Man, Cybern. {B}}, vol.~24, no.~5, pp. 769-777,
  May 1994.


\bibitem{SLA}
Y.~Xing and R.~Chandramouli, ``Stochastic learning solution for distributed
  discrete power control game in wireless data networks,'' \emph{{IEEE/ACM}
  Trans. Netw.}, vol.~16, no.~4, pp. 932-944, 2008.

  \bibitem{SLA2}
W.~Zhong, Y.~Xu, M.~Tao, et al., ``Game theoretic multimode precoding
  strategy selection for MIMO multiple access channels,'' \emph{{IEEE} Signal
  Process. Lett.}, vol.~17, no.~6, pp. 563-566, 2010.

   \bibitem{Q_learning_AAMAS10}
C.~Wu, K.~Chowdhury, M.~D.~Felice and W.~Meleis, ``Spectrum management of cognitive radio using multi-agent
reinforcement learning," Proc. of 9th Int. Conf. on Autonomous Agents and
Multiagent Systems (AAMAS 2010), pp.~1705-1712, 2010.

  \bibitem{Q_learning_JSAC13}
W.~R.~Zame, J.~Xu, and M.~van der Schaar, ``Cooperative multi-agent learning and coordination for cognitive radio networks,"
 \emph{IEEE J. Sel. Areas Commun., } vol.~32, no.~3, pp.~464-477, 2014.





\bibitem{Li_EURASIP10}
H.~Li, "Multi-agent Q-learning for Aloha-like spectrum access in cognitive radio systems,"
\emph{EURASIP Journal on Wireless Communications and Networking,} vol.~2010, pp.~1-15.

 \bibitem{Li09}
H.~Li, ``Multi-agent q-learning of channel selection in multi-user cognitive
  radio systems: A two by two case,'' in \emph{Pro. IEEE Conference on System, Man and Cybernetics (SMC)}, pp. 1893--1898, 2009

 \bibitem{Li10}
 H.~Li, ``Multi-agent q-learning for competitive spectrum access in cognitive radio systems." in
   \emph{IEEE Fifth Workshop on Networking Technologies for Software Defined Radio Networks}, 2010.

\bibitem{Learning_Cost}
M.~Khan, H.~Tembine and A.~Vasilakos, ``Game dynamics and cost of learning in heterogeneous 4G networks,"
\emph{{IEEE} J. Sel. Areas Commun.}, vol.~30, no.~1, pp.~198-213, 2012.

\bibitem{trial_learning}
L.~Rose, S.~Perlaza, C.~Martret and  M\'{e}rouane Debbah, ``Self-organization in decentralized networks: A trial and error learning approach,"
\emph{{IEEE} Trans. Wireless Commun.}, vol.~13, no.~1, pp.~268--279, 2014.


%
%



   \bibitem{effective_capacity7}
   S.~Lien,  Y.~Lin and  K.~Chen, ``Cognitive and game-theoretical radio resource management
    for autonomous femtocells with QoS guarantees," \emph{{IEEE} Trans. Wireless Commun.}, vol.~10, no.~7, pp.~2196-2206, 2011.

    \bibitem{satisfaction_equilibrium}
   S.~Perlaza,  H.~Tembine and  S.~Lasaulce, ``Quality-of-service provisioning in decentralized networks: A satisfaction equilibrium approach,"
   \emph{IEEE J. Sel. Signal Process},  vol.~6, no.~2, pp.~104-116, 2012.

   \bibitem{ICC17_Xu}
   Y.~Xu,  J.~Wang, Q.~Wu, J.~Zheng, L.~Shen and A.~Anpalagan, "Opportunistic spectrum access with statistical QoS provisioning:
   an effective-capacity based multi-agent learning approach," \emph{IEEE International
   Conference on Communications (ICC'17)}, accepted.

  \bibitem{pricing}
N.~Dusit, E.~Hossain, and Z.~Han, ``Dynamic spectrum access in IEEE 802.22-based cognitive wireless networks: A
 game theoretic model for competitive spectrum bidding and pricing,"\emph{ IEEE Wireless Communications},
  vol.~16, no.~2, pp.16--23, 2009.

\bibitem{auction}
L.~Gao, Y.~Xu, and X.~Wang, ``MAP: Multiauctioneer progressive auction for dynamic spectrum access,"
 \emph{IEEE Transactions on  Mobile Computing,} vol.~10, no.~8, pp.~1144--1161, 2011.

\bibitem{bargaining}
  J.~Suris, L.~Dasilva, Z.~Han, A.~Mackenzie, and R.~Komali, ``Asymptotic
optimality for distributed spectrum sharing using bargaining solutions,"
\emph{IEEE Trans. Wireless Commun.,} vol.~8, no.~10, pp.~5225--5237, Oct. 2009.

\bibitem{Coalitional1}
D.~Li, Y.~Xu, X.~Wang, et al, ``Coalitional game theoretic approach for secondary spectrum access in cooperative
cognitive radio networks", \emph{IEEE Trans. Wireless Commun.,} vol.~10, no.~3, pp.~844--856, 2011.


\bibitem{Coalitional2}
W.~Saad, Z.~Han, R.~Zheng, et al, ``Coalitional games in partition form for joint spectrum sensing and access in
 cognitive radio networks," \emph{IEEE J. Sel. Signal Process}, vol.~6, no.~2, pp.~195--209, 2012.

%
%
%



   \bibitem{Babadi10}
B.~Babadi, and V.~Tarokh, ``GADIA: A greedy asynchronous distributed interference avoidance algorithm,"
\emph{ IEEE Trans. Inf. Theory}, vol.~56, no.~12, pp.~6228--6252, 2010.

\bibitem{Cao08}
L.~Cao and H.~Zheng, ``Distributed rule-regulated spectrum sharing,"
\emph{{IEEE} J. Sel. Areas Commun.},, vol.~26, no.~1, pp.130--145, 2008.

\bibitem{Xu_magazine}
Y.~Xu, J.~Wang, Q.~Wu, Z.~Du, L.~Shen and A.~Anpalagan, ``A game theoretic perspective on
self-organizing optimization for cognitive small cells," \emph{IEEE Communications Magazine},
vol.~53, no.~7, pp.~100-108, 2015.

   \bibitem{finite_rate_model}
   W.~Liu,  L.~Zhou and  B.~Giannakis, ``Queuing with adaptive modulation and coding over wireless links: Cross-layer analysis and design,"
   \emph{IEEE Trans. Wireless Commun.},  vol.~4, no.~3, pp.~1142-1153, 2005.





  \bibitem{Large_deviation}
K.~B.~Letaief and J.~S.~Sadowsky, ``Computing bit error probabilities
for avalanche photodiode receivers using large deviations theory,"
\emph{IEEE Trans. Inform. Theory,} vol.~38, no.~3, pp.~1162-1169, 1992.

\bibitem{effective_capacity}
D.~Wu and R.~Negi, ``Effective capacity: A wireless link model for support of quality of service,"
\emph{{IEEE} Trans. Wireless Commun.}, vol.~2, no.~4, pp.~630-643, 2003.



\bibitem{inequality}
M.~Kuczma,\emph{ An Introduction To The Theory Of Functional Equations and Inequalities: Cauchy'S
Equation And Jensen'S Inequality}, Springer, 2008.

\bibitem{Vcking06congestiongames}
B. Vcking and R. Aachen, ``Congestion games: optimization in compe
tition," \emph{in Proc. 2006 Algorithms and Complexity in Durham Workshop,}
pp. 9-20.


\bibitem{learning_automata}
P.~Sastry, V.~Phansalkar, and M.~Thathachar, ``Decentralized learning of nash
  equilibria in multi-person stochastic games with incomplete information,''
  \emph{{IEEE} Trans. Syst., Man, Cybern. {B}}, vol.~24, no.~5, pp. 769-777,
  May 1994.

















%
%


%
%

\bibitem{HIPERLAN_2}
A.~Doufexi, S.~Armour, M.~Butler, et al., ``A comparison of the HIPERLAN/2 and IEEE 802.11 a wireless LAN standards,"
\emph{IEEE Communications Magazine}, vol.~40, no.~5, pp.~172-180, 2002.














\end{thebibliography}

\end{document}